\documentstyle[twoside,fleqn,espcrc2]{article}

\newcommand{\AmS}{{\protect\the\textfont2
  A\kern-.1667em\lower.5ex\hbox{M}\kern-.125emS}}

\title{Solving Gauss' Laws and Searching Dirac Observables
       for the Four Interactions}

\author{Luca Lusanna\address{Sezione INFN di Firenze,\\ 
        Largo E.Fermi 2 (Arcetri), 50125 Firenze, Italy \\ 
        email LUSANNA@FI.INFN.IT}}
       
\begin{document}

\begin{abstract}

A review is given of the status of the program of classical reduction to
Dirac's observables of the four interactions (standard SU(3)xSU(2)xU(1)
particle model and tetrad gravity) with the matter described either by
Grassmann-valued fermion fields or by particles with Grassmann charges.
\hfill
\hfill

Talk given at the Second Conf. on Constrained Dynamics and Quantum Gravity,
S.Margherita Ligure, 17-21 September 1996.

\end{abstract}

\maketitle

The realization that all relevant physical relativistic systems are described 
at the classical level by singular Lagrangians and Dirac-Bergmann Hamiltonian
theory of constraints \cite{di,be}, stimulated a research program trying to 
find a unification of the mathematical description of the four 
(electroweak, strong and gravitational) interactions and, after a canonical
reduction, their reformulation only in terms of Dirac's observables without any
kind of gauge degrees of freedom. See the review papers
\cite{re} for the formulation and the evolution of the program.
See Refs.\cite{lich} for the differential geometric setting behind constraint
theory (in particular the theory of presymplectic manifolds in the case of
first class constraints\cite{lich,go}) and Refs.\cite{lus} for the
reformulation of the theory of singular Lagrangians and of constraints in
terms of the second Noether theorem.
The adopted definition of Dirac's observables is based on the Shanmugadhasan
canonical tralsformation\cite{sha} in the eleboration of Refs.
\cite{lus1,lus2,re} (see also Refs.\cite{fulp})  based on the transformation
of the Hamilton-Dirac equations in the so-called multitemporal equations
\cite{re} (see Ref.\cite{bat} for previous attempts to define 
multiparametric equations describing the gauge transformations generated by
first class constraints).

For systems with only first class constraints, like those describing the four 
interactions, one looks for new canonical
bases in which all first class constraints are replaced by a subset of the new
momenta (Abelianization of the constraints); then the conjugate canonical
variables are  Abelianized gauge variables and the remaining canonical pairs
are special Dirac observables in strong involution with both Abelian
constraints and gauge  variables. These Dirac observables, together with the
Abelian gauge variables, form a local Darboux system of coordinates for the
presymplectic manifold $\bar \gamma$ defined by the original first class
constraints (this manifold is coisotropically embedded\cite{go} in the
original phase space, if suitable mathematical conditions are satisfied). 
In the multi-temporal method each first class constraint
is raised to the status of a Hamiltonian with a scalar parameter
describing the associated evolution (the genuine time describes the evolution
generated by the canonical Hamiltonian, after extraction from it of the
secondary and higher order first class constraints): in the Abelianized form
of the constraints these "times" coincide with the Abelian gauge variables on
the solutions of the Hamilton equations. These coupled Hamilton equations are
the multi-temporal equations: their solution describes the dependence of the
original canonical variables on the time and on the parameters of the
infinitesimal gauge transformations, generated by the first class constraints.
Given an initial point on the constraint manifold, the general solution
describes the gauge orbit, spanned by the gauge variables, through that point;
instead the time evolution generated by the canonical Hamiltonian (a first
class quantity) maps one gauge orbit into another one. For each system the
main problems are whether the constraint set is a manifold (a stratified
manifold, a manifold with singularities...), whether the gauge orbits can be
built in the large starting from infinitesimal gauge transformations and
whether the foliation of the constraint manifold (of each stratum of it) is
either regular or singular. Once these problems are understood, one can check
whether the reduced phase space (Hamiltonian orbit space) is well defined.

Since for all isolated systems defined on Minkowski spacetime there is the
Poincar\'e kinematical symmetry group globally canonically implemented
\cite{pau} [for field theories the boundary conditions on the fields must be 
such that the ten Poincar\'e generators are finite], the presymplectic manifold 
$\bar \gamma$ is a stratified manifold with the main stratum (dense in $\bar 
\gamma$) containing all configurations belonging to timelike Poincar\'e
realizations with spin [$P^2 > 0$, $W^2=-P^2{\hat {\vec S}}^2\not= 0$; ${\hat
{\vec S}}$ is the rest-frame Thomas spin]. Then there will be strata with
i) $P^2 > 0$, $W^2=0$, and ii) $P^2=0$; iii) $P^{\mu}=0$, the infrared stratum;
the spacelike stratum $P^2 < 0$ must be
absent, otherwise there would be configurations of the system violating
Einstein causality. Each stratum may have further stratifications and/or 
singularity structures according to the nature of the physical system.
Therefore, the canonical bases best adapted to each physical system will be the
subset of Shanmugadhasan bases which, for each Poincar\'e stratum, is also
adapted to the geometry of the corresponding Poincar\'e orbits with their
little groups. These special bases could be named Poincar\'e-Shanmugadhasan
(PS) bases for the given Poincar\'e stratum of the presymplectic manifold;
till now only the main stratum $P^2 > 0$, $W^2\not= 0$, has been investigated.
Usually PS bases are defined only locally and one needs an atlas of these charts
to cover the given Poincar\'e stratum of $\bar \gamma$; for instance this
always happens with compact phase spaces.
When the main stratum of a noncompact physical system admits a set of global
PS bases (i.e. atlases with only one chart), we get a global symplectic
decoupling (a strong form of Hamiltonian reduction) of the gauge degrees of
freedom from the physical Dirac observables without introducing gauge-fixing
constraints; this means that the global PS bases give coordinatizations of the
reduced phase space (the space of Hamiltonian gauge orbits or symplectic
moduli space).

The program of symplectic decoupling was initiated by Dirac himself\cite{dir},
who found the Dirac observables of the system composed by the electromagnetic
field and by a fermionic (Grassmann-valued) field (whose Dirac observables are
fermion fields dressed with a Coulomb cloud). Subsequently global Dirac 
observables (i.e. PS bases) were found for the following systems:

a) Relativistic two-body systems with action-at-a-distance interactions
\cite{longhi}.

b) The Nambu string\cite{colomo}.

c) Yang-Mills theory with Grassmann-valued fermion fields\cite{lusa} 
in the case of a trivial principal
bundle over a fixed-$x^o$ $R^3$ slice of Minkowski spacetime with suitable
Hamiltonian-oriented boundary conditions. After a discussion of the
Hamiltonian formulation of Yang-Mills theory, of its group of gauge
transformations and of the Gribov ambiguity, the theory has been studied in
suitable  weigthed Sobolev spaces where the Gribov ambiguity is absent.
The physical Hamiltonian has been obtained: it is nonlocal but without any
kind of singularities, it has the correct Abelian limit if the structure 
constants are turned off, and it contains the explicit realization of the 
abstract Mitter-Viallet metric.

d) The Abelian and non-Abelian SU(2)
Higgs models with fermion fields\cite{lv1}, where the
symplectic decoupling is a refinement of the concept of unitary gauge.
There is an ambiguity in the solutions of the Gauss law constraints, which
reflects the existence of disjoint sectors of solutions of the Euler-Lagrange
equations of Higgs models. The physical Hamiltonian and Lagrangian of  the
Higgs phase have been found; the self-energy turns out to be local and
contains a local four-fermion interaction. 

e) The standard SU(3)xSU(2)xU(1) model of elementary particles\cite{lv3}.
The final reduced Hamiltonian contains nonlocal self-energies for the
electromagnetic and color interactions, but ``local ones" for the weak 
interactions implying the nonperturbative emergence of 4-fermions interactions.
 To obtain a nonlocal self-energy with a Yukawa kernel for the
massive Z and $W^{\pm}$ bosons one has to reformulate the model on spacelike
hypersurfaces and make a modification of the Lagrangian.

Moreover, also nonrelativistic Newton mechanics has been reformulated as a
many-times theory with first class constraints\cite{nm}. With a suitable
contraction on the ADM Lagrangian for general relativity, a singular
Lagrangian for Newtonian gravity with general Galilei covariance has been found
\cite{ng}: there are 27 fields and many first and second class constraints at 
the Hamiltonian level connected with the inertial forces and the Newton
gravitational potential.

However, all the Hamiltonian reductions of relativistic gauge field theories 
suffer of the problem of Lorentz covariance: one cannot make a complete
Hamiltonian reduction for systems defined in Minkowski spacetime without a
breaking of manifest Lorentz covariance. A universal solution of this problem
has been found by reformulating\cite{lusa,lus3,lus4}  every relativistic system
on a family of spacelike hypersurfaces foliating Minkowski spacetime\cite{di}.
In Ref.\cite{lus3} there is  a consistent formultation of the system of
N scalar charged particles (with the charges described in a pseudoclassical way
with Grassmann variables\cite{casal})
and of the electromagnetic field on spacelike
hypersurfaces, with the Coulomb potential extracted from field theory by means
of the canonical reduction to Dirac's observables (for both the field and the
particles, which are dressed with a Coulomb cloud) and with the pseudoclassical
self-energies regularized by the Grassmann charges. 
On spacelike hypersurfaces, one has a covariant 1-time description of scalar
particles, in which each particle is described by 3 coordinates on the
hypersurface. This implies that the mass-shell constraint for each particle
has been solved and that one has done a choice of the sign of the energy of
each particle. This solves the covariance problem present in the standard
manifestly covariant Lagrangian description of charged particles plus the
electromagnetic field: at the Hamiltonian level one does not know how to
eveluate the Poisson bracket of the particle mass-shell primary constraints
with the electromagnetic primary constraints, because there is no notion of
equal time among them. The extension of this 
approach to colored particles (with the color described by Grassmann variables)
and with Yang-Mills theory reformulated on spacelike hypersurfaces
is beginning to produce the pseudoclassical basis of the
relativistic quark model\cite{alba}: between two color charge densities (either
a particle with color charge or field charge density) three kinds of 
interactions act [direct Coulomb interaction; direct Coulomb interaction
between one density and a point (over whose location it is integrated) and a
Dirac-observable Wilson line interaction along the flar geodesic between the
point and the other density; each density interacts with a point through a
geodesic Wilson line and there is a Coulomb interaction between the two points
(over whose locations it is integrated)] and
there already is an indication of the
presence of asymptotic freedom in the quark-antiquark sector under the
hypothesis of color singlets (zero quark-antiquark total color charge plus no
flux at infinity of total Yang-Mills non-Abelian color charge).

Therefore, at this stage one knows the description on spacelike hypersurfaces
of scalar particles and of bosonic fields (see Ref.\cite{ku} for the
general theory).
What is still to be done is the reformulation of Grassmann-valued Dirac fields 
on spacelike hypersurfaces, which is connected with the treatment of tetrad
gravity and the description of spinning particles.

In this way one obtains the minimal breaking of 
Lorentz covariance: after the restriction from arbitrary spacelike
hypersurfaces to spacelike hyperplanes, one selects all the configurations
belonging to the main Poincar\'e stratum and then restricts oneself to
the special family of hyperplanes orthogonal to the total momentum of the
given configuration (this family may be called the Wigner foliation of
Minkowski spacetime intrinsically defined by the given system).
In this way only three physical degrees of freedom,
describing the canonical center-of-mass 3-position of the overall isolated
system, break Lorentz covariance, while all the field variables are either
Lorentz scalars or Wigner spin-1 3-vectors transforming under Wigner
rotations. This method is based on canonical realizations of the Poincar\'e
group on spaces of functions on phase spaces\cite{pau} 
and one obtains the ``covariant 1-time rest-frame instant form" of the dynamics
and also the covariant 1-time relativistic statistical mechanics\cite{lus3}.

Therefore one has to study the problem of the center of mass of
extended relativistic systems in irreducible representations of the Poincar\'e
group with $P^2 > 0$, $W^2=-P^2{\vec {\bar S}}^2\not= 0$ : it can be shown
that this problem leads to the identification of a finite world-tube of
non-covariance of the canonical center-of-mass, whose radius $\rho =\sqrt{-W^2}
/P^2=|\, {\vec {\bar S}}\, |\, /\sqrt{P^2}$ \cite{mol}
 identifies a classical intrinsic
unit of length, which can be used as a ultraviolet cutoff at the quantum
level in the spirit of Dirac and Yukawa.
As mentioned in the papers\cite{lusa,lus3}, the distances
corresponding to the interior of the world-tube are connected with problems
coming from both quantum theory and general relativity: 1) pair production 
happens when trying to localize particles at these distances; 2) relativistic
bodies with a material radius less than $\rho$ cannot have the classical
energy density definite positive everywhere in every reference frame and the
peripheral rotation velocity may be higher than the velocity of light.
Therefore, the world-tube also is the flat remnant of the energy  conditions of
general relativity. A problem under investigation is the definition of the
center of mass of classical fields, because
this is the classical background of Tomonaga-Schwinger formulation 
of the relativistic Cauchy problem for quantum field theory\cite{lon}.

The next step of the program is the search of Dirac's observables for classical
tetrad gravity\cite{rus} [with asymptotically flat spacetimes so to have the 
asymptotic Poincar\'e charges\cite{poi} 
and the same ultraviolet cutoff $\rho$ as for
the other interactions], which replaces metric gravity when couplings to
fermion fields have to be done, and its coupling to the standard particle model.

Since the previous approaches to tetrad gravity\cite{tetrad} were all using
in some form Schwinger's time gauge condition, a new formulation was looked for.

As for Yang-Mills theory, one has to define the simplest family of spacetimes
to have all possible differential geometry techniques available. In the
case of tetrad gravity, one assumes that the spacetime $M^4$ is globally
hyperbolic and asymptotically flat, i.e. $M^4=\Sigma \times R$, and that the
noncompact globally parallelizable 3-surface $\Sigma$, on which the
spacelike slices $\Sigma_{\tau}$ are modeled, is topologically trivial and
that the Riemannian 3-manifolds $(\Sigma_{\tau}, {}^3g)$ are 
geodesically complete, satisfying the Hopf-Rinow theorem (so that the geodesic
exponential map is defined on all the tangent space $T_p\Sigma_{\tau}$ of at
least one point $p\in \Sigma_{\tau}$) and finally satisfying the Hadamard
theorem\cite{oneil}: 
in conclusion $\Sigma$ is diffeomorphic to $R^3$, the geodesic 
exponential map is a diffeomorphism from $T\Sigma$ to $\Sigma$ in each point,
and there are no conjugate Jacobi points and no closed geodesics in $\Sigma$.
In particular, the orthonormal frame principal SO(3)-bundle over $\Sigma_{\tau}$
(whose connections are the spin connections determined by the triads) is trivial
and the 3-manifold $\Sigma_{\tau}$ admits smooth global charts.

Then, by using $\Sigma_{\tau}$-adapted coordinates, one has found a
parametrization of arbitrary tetrads and cotetrads on $M^4$ in terms of
special $\Sigma_{\tau}$-adapted tetrads and cotetrads and of 3 angles
parametrizing point-dependent Wigner boosts for timelike Poincar\'e orbits.
Then these variables are reexpressed in terms of lapse and shift functions,
of cotriads on $\Sigma_{\tau}$ and of the 3 boost angles. Putting
these variables in the ADM action for metric gravity \cite{adm}
(with the 3-metric on
$\Sigma_{\tau}$ expressed in terms of cotriads), one gets a new action
depending only on lapse, shifts and cotriads, but not on the boost angles
(therefore, there is no need of Schwinger's time gauge: the 3 primary
constraints describing the Lorentz boost on $T\Sigma_{\tau}$ are automatically
Abelianized, being described by three vanishing momenta). There are 10
primary and 4 secondary first class constraints and a vanishing canonical
Hamiltonian. Besides the 3 constraints associated with Lorentz boosts, there are
4 constraints saying that the momenta associated with lapse and shifts vanish,
3 constraints describing rotations, 3 constraints generating $\Sigma_{\tau}$
space-diffeomorphisms (a linear combination of supermomentum constraints and
of the rotation ones) and one superhamiltonian constraint. It turns out that 
with the technology developed for Yang-Mills theory, one can Abelianize the 3 
rotation constraints and then also the space-diffeomorphism constraints. In
the resulting quasi-Shanmugadhasan canonical basis, one is left in the
physical sector with reduced cotriads depending only on 3 degrees of freedom 
(they are Dirac's observables with respect to 13 of the 14 first class 
constraints) and with their conjugate momenta still subject to the reduced 
form of the superhamiltonian constrain (which now has zero Poisson bracket 
with itself): this is the phase space over the superspace of 3-geometries
\cite{witt}.

In the Abelianization of the rotation constraints one needs the Green function
of the 3-dimensional covariant derivative containing the spin connection. It
will be well defined only if there is no Gribov ambiguity and this will
require the use of weighted Sobolev spaces like in Yang-Mills theory:now there
is also the implication that isometries of the Riemannian 3-manifold
$(\Sigma_{\tau},{}^3g)$ must be absent\cite{choq}. The Green function is similar
to the Yang-Mills one for a principal SO(3)-bundle \cite{lusa}, but, instead
of the Dirac distribution\cite{dir} for the Green function of the flat
divergence, it contains the DeWitt function\cite{dew} (which reduces to the 
Dirac distribution only locally in normal coordinates). Moreover, the definition
of the Green function now requires the geodesic exponential map.
Not having a group
theoretical control on the diffeomorphism group, the Abelianization of the
space-diffeomorphism constraints is only a property of the algebra of the
infinitesimal diffeomorphisms, without a real understanding of the global
problems of finite diffeomorphisms. However the multitemporal equations
generated by the space-diffeomorphisms can be written explicetly and a
global solution can be found. The original cotriad can be expressed in closed
form in terms of 3 rotation angles, 3 diffeomorphism-parameters and the
reduced cotriad depending only on 3 independent variables.

Till now no coordinate condition\cite{ish} has been imposed. It turns out that
these conditions are hidden in the choice of how to parametrize the reduced
Dirac-observable cotriads in terms of three independent functions. The
simplest parametrization (the only one studied till now) correspond to choose a
system of 3-orthogonal coordinates on $\Sigma_{\tau}$, in which the 3-metric is
diagonal. With a further canonical transformation on the reduced cotriads and 
conjugate momenta, one arrives at a canonical basis containing the conformal
factor of the 3-geometry and its conjugate momentum plus two other pairs of
conjugate canonical variables. The reduced superhamiltonian constraint,
expressed in terms of these variables, turns out to be an integral equation
for the momentum conjugate to the conformal factor (which, therefore, is a gauge
variable) due to the signature of the DeWitt supermetric.

This is a schematic description of the preliminary results obtained in pure
tetrad gravity. A lot of work has still to be done: i) a comparison with the
standard York reduction\cite{york,choq,ciuf} based on conformal techniques, in 
which the superhamiltonian constraint becomes the elliptic Lichnerowicz
equation for the conformal factor and in which one uses York's cosmic time
\cite{quadir} (the trace of the extrinsic curvature of $\Sigma_{\tau}$, usually
assumed to be a maximal slice); ii) a better understanding of the integral 
equation; iii) a study of the Hamiltonian boundary conditions and of the
surface terms needed to have well defined Poisson brackets and Hamilton 
equations\cite{witt}; iv) a detailed study of the Hamiltonian asymptotic 
symmetries at spatial infinity\cite{poi} [see Refs.\cite{ash} for the covariant
4-dimensional spi approach to the problem] and of the boost problem\cite{chris},
to arrive at a definition of the asymptotic Poincar\'e charges (without 
supertranslation ambiguities), of the intrinsic spin and of the asymptotic
Poincar\'e Casimirs, needed to build the intrinsic unit of lenght $\rho$
(absent in a compact universe, liked by Einstein for avoiding boundary 
conditions and by Wheeler\cite{ciuf} to have a full implementation of Mach's 
ideas), which is an intrinsic candidate for a ultraviolet cutoff; v) to
verify the positivity of energy; vi) to understand the nature of the ADM slices
$\Sigma_{\tau}$ selected by this canonical reduction (replacing the maximal
slices of York's approach): for the pure gauge case (flat Minkowski spacetime
solution) the slices are conformal to spacelike hyperplanes, which are 
deformed in presence of curvature (since the $\Sigma_{\tau}$'s are not compact,
Yamabe theorem does not apply).

Further problems are how to deparametrize the theory\cite{isha}, so to
reexpress it in the form of parametrized field theories on spacelike
hypersurfaces in Minkowski spacetime. This is an extremely important point,
because, if we add N scalar particles to tetrad gravity (whose reduction to
Dirac's observables should define the N-body problem in general relativity),
the deparametrization should be the bridge to the previously quoted theory on
spacelike hypersurfaces in Minkowski spacetime\cite{lus3} in the limit of
zero curvature. A new formulation of the N-body problem would be relevant to
try to understand the energy balance in the emission of gravitational waves
from systems like binaries. If it will be possible to find the Dirac
observables for the particles, on will understand how to extract from the
field theory the covariantization of Newton potential (one expects one scalar
and one vector potential\cite{ciuf}) and a mayor problem will be how to face
the expected singularities of the mass-self-energies.

Finally one should couple tetrad gravity to the
electromagnetic field, to fermion fields and then to the standard model,
trying to make to reduction to Dirac's observables in all these cases.

Only after the completion of the classical reduction program one will start
with an attempt to quantize the nonlocal and nonpolynomial theories obtained
in this way, whose Hamiltonians, however, 
are always quadratic in the momenta and
which will have Wigner covariance when formulated in the covariant 1-time
rest-frame instant form of dynamics and in what will be, if any, its extension
to general relativity. In any case the center-of-mass variable of the isolated
system, which breaks covariance, does not need to be quantized: like the
wave function of the universe no one can observe it (it is on the classical side
of the Copenhagen interpretation). Its noncovariance gives us an intrinsic
classical unit of lenght to be used as a ultraviolet cutoff in quantization,
and the resulting picture would be some kind of weak Mach principle: only
relative motions are relavant and have to be quantized.


\begin{thebibliography}{99}





\bibitem{di}P.A.M.Dirac, Can.J.Math. {\bf 2}, 129 (1950); "Lectures on 
     Quantum Mechanics", Belfer Graduate School of Science, Monographs Series 
     (Yeshiva University, New York, N.Y., 1964).
\bibitem{be}J.L.Anderson and P.G.Bergmann, Phys.Rev. {\bf 83}, 1018 (1951).
     P.G.Bergmann and J.Goldberg, Phys.Rev. {\bf 98}, 531 (1955).
\bibitem{re}L.Lusanna, "Multitemporal
     Relativistic Particle Mechanics: a Gauge Theory without Gauge-Fixings",
     in Proc. IV Marcel Grossmann Meeting on General Relativity, ed. R.Ruffini
     (Elsevier, Amsterdam, 1986);
     "Classical Observables of Gauge Theories from the
     Multitemporal Approach", talk given at the Conference 'Mathematical
     Aspects of Classical Field Theory', Seattle 1991, in Contemporary
     Mathematics {\bf 132}, 531 (1992);
"Dirac's Observables: from Particles to Strings and 
Fields", talk given at the International Symposium on 'Extended Objects and
Bound States', Karuizawa 1992, eds. O.Hara, S.Ishida and S.Naka,
(World Scientific, Singapore, 1993);
``Hamiltonian Constraint's and Dirac's 
Observables: from Relativistic Particles towards Field Theory and General 
Relativity", talk at the Workshop ``Geometry of Constrained Dynamical Systems",
Cambridge, 1995);
``Aspects of Galilean and Relativistic Particle Mechanics
with Dirac's Constraints", talk at the Conf. ``Theories of Fundamental 
Interactions", Maynooth, Ireland, 1995, ed.D.H.Tchrakian (World Scientific,
Singapore, 1996);
``Classical Dirac Observables: the Emergence of
Rest-Frame Particle and Field Theories, talk at the Pacific Conference on
Gravitation and Cosmology, Seoul 1996.
\bibitem{lich}A.Lichnerowicz, C.R.Acad.Sci.Paris, Ser. A, {\bf 280}, 523 (1975).
W.Tulczyiew, Symposia Math. {\bf 14}, 247 (1974). J.\`Sniatycki, Ann.Inst.
H.Poincar\`e {\bf 20}, 365 (1984). G.Marmo, N.Mukunda and J.Samuel, Riv.Nuovo 
Cimento {\bf 6}, 1 (1983). M.J.Bergvelt and E.A.De Kerf, Physica {\bf 139A}, 
101 and 125 (1986). B.A.Dubrovin, M.Giordano, G.Marmo and A.Simoni,
Int.J.Mod.Phys. {\bf 8}, 4055 (1993).
\bibitem{go}M.J.Gotay, J.M.Nester and G.Hinds, J.Math.Phys. {\bf 19}, 2388 
(1978). M.J.Gotay and J.M.Nester, Ann.Inst.Henri Poincar$\grave e$ {\bf A30}, 
129 (1979) and {\bf A32}, 1 (1980). M.J.Gotay and J.$\grave S$niatycki, Commun.
Math.Phys. {\bf 82}, 377 (1981). M.J.Gotay, Proc.Am.Math.Soc. {\bf 84}, 111 
(1982); J.Math.Phys. {\bf 27}, 2051 (1986).
\bibitem{lus}L.Lusanna, Nuovo Cimento {\bf B52}, 141 (1979);
Phys.Rep. {\bf 185}, 1 (1990); Riv. Nuovo Cimento {\bf 14}, n.3, 1 (1991).
M.Chaichian, D.Louis Martinez and L.Lusanna, Ann.Phys.(N.Y.)
     {\bf 232}, 40 (1994).
\bibitem{sha}S.Shanmugadhasan, J.Math.Phys. {\bf 14}, 677 (1973).
\bibitem{lus1}L.Lusanna, Int.J.Mod.Phys. {\bf A8}, 4193 (1993).
\bibitem{lus2}L.Lusanna, J.Math.Phys. {\bf 31}, 2126 (1990); J.Math.Phys. 
{\bf 31}, 428 (1990).
\bibitem{fulp}R.O.Fulp and J.A.Marlin, Pacific J.Math. {\bf 67}, 373 (1976); 
Rep.Math.Phys. {\bf 18}, 295 (1980).
\bibitem{bat}I.A.Batalin and G.A.Vilkovisky, Nucl.Phys. {\bf B234}, 106 (1984).
\bibitem{pau}M.Pauri and G.M.Prosperi, J.Math.Phys. {\bf 7}, 366 (1966);
{\bf 16}, 1503 (1975).
\bibitem{dir}P.A.M.Dirac, Can.J.Phys. {\bf 33}, 650 (1955).
\bibitem{longhi}G.Longhi and L.Lusanna, Phys.Rev. {\bf D34}, 3707 (1986).
\bibitem{colomo}F.Colomo and L.Lusanna, Int.J.Mod.Phys. {\bf A7}, 
     1705 and 4107 (1992). F.Colomo, G.Longhi and L.Lusanna, Int.J.Mod.Phys.
{\bf A5}, 3347 (1990); Mod.Phys,Lett. {\bf A5}, 17 (1990).
\bibitem{lusa}L.Lusanna, Int.J.Mod.Phys. {\bf A10}, 3531 and 3675 (1995).
\bibitem{lv1}L.Lusanna and P.Valtancoli, ``Dirac's Observables for the Higgs
model: I) the Abelian Case; II) the non-Abelian SU(2) Case", Firenze University 
preprints, 1996 (SISSA, hep-th 9606078 and 9606079).
\bibitem{lv3}L.Lusanna and P.Valtancoli, Dirac's Observables for the
SU(3)xSU(2)xU(1) Standard Model, Firenze preprint 1996.
\bibitem{nm}G.Longhi, L.Lusanna and J.M.Pons, J.Math.Phys. {\bf 30}, 1893 
(1989). R.De Pietri, L.Lusanna and M.Pauri, Class.Quantum Grav. {\bf 13}, 
1417 (1996).
\bibitem{ng}R.De Pietri, L.Lusanna and M.Pauri, Class.Quantum Grav. {\bf 12},
     219 and 255 (1995).
\bibitem{lus3}L.Lusanna, ``N- and 1-time Classical Description of
N-body Relativistic Kinematics and the Electromagnetic Interaction",
to appear in Int.J.Mod.Phys. A (SISSA, HEP-TH/9512070).
\bibitem{lus4}L.Lusanna, in Gauge Field Theories, XVIII Karpacz School, 1981, 
ed. W.Garczynski (Harwood, Chur, 1986).
\bibitem{casal}R.Casalbuoni, Nuovo Cimento {\bf A33}, 115, 389 (1976). F.A.
Berezin and M.S.Marinov, Ann.Phys.(N.Y.) {\bf 104}, 336 (1977). A.Barducci, 
R.Casalbuoni and L.Lusanna, Nuovo Cimento {\bf A35}, 377 (1976); Nuovo Cimento
Lett. {\bf 19}, 581 (1977); Nucl.Phys. {\bf B124}, 93 (1977); Nucl.Phys.
{\bf B180} [FS2], 141 (1981).
\bibitem{alba}D.Alba and L.Lusanna, in preparation.
\bibitem{ku}K.Kuchar, J.Math.Phys. {\bf 17}, 777, 792, 801 (1976); {\bf 18},
1589 (1977).
\bibitem{mol}C.M$\o$ller, "The Theory of Relativity" (Oxford Univ.Press, 
Oxford, 1957). M.Pauri, "Invariant Localization and Mass-Spin Relations in the
     Hamiltonian Formulation of Classical Relativistic Dynamics", Parma Univ.
     preprint IFPR-T-019, 1971 (unpublished); in "Group Theoretical Methods
     in Physics", ed. K.B.Wolf, Lecture Notes Phys. n.135 (Springer, Berlin,
     1980).
\bibitem{lon}G.Longhi and M.Materassi, in preparation
\bibitem{rus}L.Lusanna and S.Russo, in preparation.
\bibitem{poi}T.Regge and C.Teitelboim, Ann.Phys.(N.Y.) {\bf 88}, 286 (1974).
R.Beig and N.\'O Murchadha, Ann.Phys.(N.Y.) {\bf 174}, 463
(1987).
L.Andersson, J.Geom.Phys. {\bf 4}, 289 (1987).
\bibitem{tetrad}H.Weyl, Z.Physik {\bf 56}, 330 (1929).
P.A.M.Dirac, in ``Recent Developments in General Relativity",
(Pergamon Press, Oxford, and PWN-Polish Scientific Publishers,
Warsaw, 1962). J.Schwinger, Phys.Rev. {\bf 130}, 1253 (1963).
T.W.B.Kibble, J.Math.Phys. {\bf 4}, 1433 (1963).
S.Deser and C.J.Isham, Phys.Rev. {\bf D14}, 2505 (1976).
J.E.Nelson and C.Teitelboim, Ann.Phys.(N.Y.) {\bf 116}, 86 (1978).
M.Pilati, Nucl.Phys. {\bf B132}, 138 (1978).
L.Castellani, P.van Nieuwenhuizen and M.Pilati, Phys.Rev. {\bf D26}, 352 (1982).
J.E.Nelson and T.Regge, Ann.Phys.(N.Y.) {\bf 166}, 234 (1986); Int.J.Mod.Phys.
{\bf A4}, 2021 (1989). J.M.Charap and J.E.Nelson, J.Phys. {\bf A16}, 1661 and 
3355 (1983). Class.Quantum Grav. {\bf 3}, 1061 (1986).
J.M.Charap, The Constraints in Vierbein General Relativity, in Constraint's
Theory and Relativistic Dynamics, eds. G.Longhi and L.Lusanna (World
Scientific, Singapore, 1987). J.W.Maluf, Class.Quantum Grav. {\bf 8}, 287 
(1991). M.Henneaux, Phys.Rev. {\bf D27}, 986 (1983).
M.Charap, M.Henneaux and J.E.Nelson, Class.Quantum Grav. 
{\bf 5}, 1405 (1988).
M.Henneaux, J.E.Nelson and C.Schonblond, Phys.Rev. {\bf D39},
434 (1989).
\bibitem{oneil}B.O'Neil, Semi-Riemannian Geometry (Academic Press, New York,
1983).
\bibitem{adm}B.S.De Witt, Phys.Rev. {\bf 160}, 1113 (1967).
R.Arnowitt, S.Deser and C.W.Misner, Phys.Rev. {\bf 117}, 1595
(1960); in Gravitation: an Introduction to Current Research, ed.L.Witten
(Wiley, New York, 1962).
C.J.Isham, ``Canonical Quantum Gravity and the Problem of Time",
in ``Integrable Systems, Quantum Groups and Quantum Field Theories",
eds.L.A.Ibort and M.A.Rodriguez, Salamanca 1993 (Kluwer, London, 1993);
``Conceptual and Geometrical Problems in Quantum Gravity", in ``Recent Aspects
of Quantum Fields", Schladming 1991, eds. H.Mitter and H.Gausterer 
(Springer, Berlin, 1991); ``Prima Facie Questions in Quantum Gravity" and 
``Canonical Quantum Gravity and the Question of Time", in ``Canonical Gravity:
From Classical to Quantum", eds. J.Ehlers and H.Friedrich (Springer, Berlin,
1994).
K.Kuchar, Time and Interpretations of Quantum Gravity, in Proc.
4th Canadian Conf. on General Relativity and Relativistic Astrophysics,
eds. G.Kunstatter, D.Vincent and J.Williams (World Scientific, Singapore,
1992). J.B.Romano, Gen.Rel.Grav. {\bf 25}, 759 (1993).
R.Beig, ``The Classical Theory of Canonical General Relativity",
in ``Canonical Gravity: From Classical to Quantum",
Bad Honnef 1993, eds. J.Ehlers and H.Friedrich, Lecture Notes Phys. 434 
(Springer, Berlin, 1994).
\bibitem{witt}B.S.De Witt, Phys.Rev. {\bf 160}, 1113 (1967).
\bibitem{choq}Y.Choquet-Bruhat and J.W.York jr., ``The Cauchy Problem", in
``General Relativity and Gravitation", Vol.1, ed.A.Held (Plenum, New York,
1980).
\bibitem{dew}B.S.De Witt, Phys.Rev. {\bf 162}, 1195 (1967); The Dynamical Theory
of Groups and Fields (Gordon and Breach, New York, 1967) and in Relativity,
Groups and Topology, Les Houches 1963, eds. C.De Witt and B.S.De Witt (Gordon
and Breach, London, 1964); The Spacetime
Approach to Quantum Field Theory, in Relativity, Groups and Topology II,
Les Houches 1983, eds. B.S.DeWitt and R.Stora (North-Holland, Amsterdam, 1984).
\bibitem{ish}P.A.M.Dirac, Phys.Rev. {\bf 114}, 924 (1959).
C.J.Isham and K.Kuchar, Ann.Phys.(N.Y.) {\bf 164}, 288 and 316
(1984). K.Kuchar, Found.Phys. {\bf 16}, 193 (1986).
\bibitem{york}J.W.York jr, Phys.Rev.Lett. {\bf 26}, 1036 (1971); {\bf 28},
1082 (1972). J.Math.Phys. {\bf 13}, 125 (1972); {\bf 14}, 456 (1972).
Ann.Ins.H.Poincar\'e {\bf XXI}, 318 (1974).
N.O'Murchadha and J.W.York jr, J.Math.Phys. {\bf 14}, 1551 (1972). Phys.Rev.
{\bf D10}, 428 (1974).
J.W.York jr., ``Kinematics and Dynamics of General Relativity",
in ``Sources of Gravitational Radiation", Battelle-Seattle Workshop 1978,
ed.L.L.Smarr (Cambridge Univ.Press, Cambridge, 1979).
\bibitem{ciuf}I.Ciufolini and J.A.Wheeler, Gravitation and Inertia (Princeton
Univ.Press, Princeton, 1995).
\bibitem{quadir}A.Qadir and J.A.Wheeler, ``York's Cosmic Time Versus Proper
Time", in ``From SU(3) to Gravity", Y.Ne'eman's festschrift, eds. E.Gotsma
and G.Tauber (Cambridge Univ.Press, Cambridge, 1985).
\bibitem{ash}A.Ashtekar, ``Asymptotic Structure of the Gravitational Field at
Spatial Infinity", in ``General Relativity and Gravitation", Vol. 2, ed.A.Held
(Plenum, New York, 1980).
A.Ashtekar, ``On the Boundary Conditions for Gravitational and
Gauge Fields at Spatial Infinity", in ``Asymptotic Behaviour of Mass and 
Spacetime Geometry" ed.F.J.Flaherty, Lecture Notes Phys.202 (Springer, Berlin,
1984).
A.Ashtekar and J.D.Romano, Class.Quantum Grav. {\bf 9}, 1069
(1992).
\bibitem{chris}D.Christodoulou and N.\'O Murchadha, Commun.Math.Phys. {\bf 80}, 
271 (1981).
\bibitem{isha}C.J.Isham and K.Kuchar, Ann.Phys.(N.Y.) {\bf 164}, 288 and 316
(1984). K.Kuchar, Found.Phys. {\bf 16}, 193 (1986).




\end{thebibliography}
\end{document}